\documentclass[10pt, conference, compsocconf]{IEEEtran}
\IEEEoverridecommandlockouts

\usepackage[toc,acronym]{glossaries}
\usepackage{adjustbox}
\usepackage{tikz}
\usetikzlibrary{arrows,shadows,backgrounds,calc,decorations.pathreplacing,decorations.pathmorphing,positioning,
	automata}
\usepackage{pgfplots}
\pgfplotsset{compat=1.13}
\usepackage{pgfplotstable}
\makeglossaries

\newacronym{AI}{AI}{Artificial Intelligence}
\newacronym{ALU}{ALU}{Arithmetical and Logical Unit}
\newacronym{CPU}{CPU}{Central Processing Unit}
\newacronym{FPGA}{FPGA}{Field Programmable Gate Array}
\newacronym{HW}{HW}{hardware}
\newacronym{ISA}{ISA}{Instruction Set Architecture}
\newacronym{I/O}{I/O}{Input/Output}
\newacronym{LAN}{LAN}{Local Area Network}
\newacronym{MC}{MC}{Multi-Core and/or Many-Core}
\newacronym{MLP}{MLP}{Memory Level Parallelism}
\newacronym{OoO}{OoO}{Out-of-Order}
\newacronym{OS}{OS}{operating system}
\newacronym{PD}{PD}{Propagation Delay}
\newacronym{QT}{QT}{Quasi-Thread}
\newacronym{PU}{PU}{Processing Unit}
\newacronym{SPA}{SPA}{Single Processor Approach}
\newacronym{SW}{SW}{software}
\newacronym{HPL}{HPL}{High Performance Linpack}
\newacronym{HPCG}{HPCG}{High Performance Conjugate Gradients}

\newacronym{EMPA}{EMPA}{Explicitly Many-Processor Approach}
\newacronym{EPE}{EPE}{EMPA Processing Element}
\newacronym{EME}{EME}{EMPA Morphing Element}
\newacronym{ECE}{ECE}{EMPA Communicating Element}
\newacronym{EICB}{EICB}{EMPA Inter-Core Block}
\newacronym{ESME}{ESME}{EMPA Storage Manager Element}

\definecolor{webgreen}{rgb}{0,.5,0}
\definecolor{webbrown}{rgb}{.6,0,0}
\definecolor{webyellow}{rgb}{0.98,0.92,0.73}
\definecolor{webgray}{rgb}{.753,.753,.753}
\definecolor{webblue}{rgb}{0,0,.8}
\definecolor{webgreen}{rgb}{0, 0.5, 0} 
\definecolor{webred}{rgb}{0.5, 0, 0}   

\usepackage{cite}
\usepackage{amsmath,amssymb,amsfonts}
\usepackage{graphicx}
\usepackage{textcomp}
\usepackage{xcolor}
\def\BibTeX{{\rm B\kern-.05em{\sc i\kern-.025em b}\kern-.08em
    T\kern-.1667em\lower.7ex\hbox{E}\kern-.125emX}}

\begin{document}

\title{How to extend the Single-Processor Paradigm\\
	to the Explicitly Many-Processor Approach\\
\thanks{Projects no. 125547 has been implemented with the support provided from the National Research, Development and Innovation Fund of Hungary, financed under the K funding scheme.
		}
\thanks{Submited to 2020 International Conference on Computational Science and Computational Intelligence (CSCE), Las Vegas, US
}
}

\author{\IEEEauthorblockN{J\'anos  V\'egh}
\IEEEauthorblockA{\textit{Kalim\'anos BT} \\
Debrecen, Hungary \\
Vegh.Janos@gmail.com~ORCID:~0000-0002-3247-7810}
}

\maketitle

\begin{abstract}
The computing paradigm invented for processing a small amount of data on a single segregated processor cannot meet the challenges set by the present-day computing demands.
The paper proposes a  new computing paradigm (extending the old one to use several processors explicitly)
and discusses some questions of its possible implementation. Some advantages of the implemented approach, illustrated with the results of a loosely-timed simulator, are presented.
\end{abstract}

\begin{IEEEkeywords}
modern computing paradigm, performance limitation, efficiency, parallelized computing,  distributed computing
\end{IEEEkeywords}

%
%
%
\section{Introduction}\label{sec:Introduction}

Physical implementations of the 70-year old computing paradigm has several limitations~\cite{LimitsOfLimits2014}. As the time passes, more and more difficulties come to light, but development of processor, the \textit{central} element of a computer,
could keep pace with the growing demand on computing till some point.
Around 2005 it became evident that the price paid for keeping the \gls{SPA} paradigm~\cite{AmdahlSingleProcessor67}, (as Amdahl coined the wording), became too high. "The   \textit{implicit   hardware/software contract} was, that increases in transistor count and power dissipation were  OK  as long as architects maintained the existing sequential programming model.  This  contract  led  to  innovations  that  were  inefficient  in transistors  and  power -- such  as  multiple  instruction  issue,  deep pipelines,   out-of-order   execution,   speculative   execution,   and prefetching -- but  which  increased  performance  while  preserving the sequential programming model."~\cite{AsanovicParallelCACM:2009}
The conclusion was that
"\textit{new ways of exploiting the silicon real estate need to be explored}"~\cite{ChandyParallelism:2009}.

"\textit{Future growth in computing performance must come
from parallelism}"~\cite{ComputingPerformance:2011} is the common point of view. However, "\textit{when we start talking about parallelism and
ease of use of truly parallel computers, we’re
talking about a problem that’s as hard as any that
computer science has faced}."~\cite{AsanovicParallelCACM:2009}.
Mainly because of this, parallel utilization of computers could not replace the energy-wasting solutions introduced to the formerly favorited
single-thread processors. They remained in the \gls{MC} processors, greatly contributing to their dissipation and, through this, to the overall crisis of computing~\cite{GameOverYelick:2011}.

Computing paradigm itself, the \textit{implicit   hardware/software  contract}, was suspected
even more explicitly:
"\emph{Processor and network architectures are making rapid progress with more and more cores being integrated into single processors and more and more machines getting connected with increasing bandwidth. Processors become heterogeneous and reconfigurable
	\dots No current programming model is able to cope with this development, though, as they essentially still follow the classical van Neumann model.}"~\cite{SoOS:2010}
On one side, 
when thinking about "advances beyond 2020", the solution was expected from the "\textit{more efficient implementation of the von Neumann architecture}"~\cite{DeBenedictis_supercomputing:2014}. On the other side, there are statements such as "\textit{The von Neumann architecture} is fundamentally inefficient
and non-scalable for representing massively interconnected neural networks"~\cite{TrueNorth:2016}.

It is worth, therefore, to scrutinize
that \textit{implicit  hardware/software   contract}, whether the processor architecture could be better adapted to the changes that occurred in the past seven decades in the technology and utilization of computing. \textit{Implicitly}, both the \gls{HW} and \gls{SW} solutions advantageously use multi-processing.
The paper shows that using a less rigid interpretation of the terms that that contract is based upon, one can extend the single-thread paradigm to use several processors \textit{explicitly} (enebling direct core-to-core interaction), without violating the 'contract', the 70-year old \gls{HW}/\gls{SW} interface.

Section \ref{sec:EMPAgeneral} shortly summarizes some of the major challenges the modern computing is expected to cope with and sketches the principles that enable it to give a proper reply.
The way to implement those uncommon principles proposed here is discussed in section~\ref{sec:Implementation}.
Because of the limited space, only a few of the advantages are demonstrated in section~\ref{sec:NewFeatures}.

\section{The general principles of EMPA\label{sec:EMPAgeneral}}

During the past two decades, computing developed in direction to conquer also some extremes: the 'ubiquitous computing' led to billions of connected and interacting processors~\cite{FutureIoT2013}, the always higher need for more/finer details, more data and shorter processing times led to building computers comprising millions of processors to target challenging tasks~\cite{Scienceexascale:2018}, different cooperative solutions~\cite{FogCloudComputing:2018} attempt to handle the demand of dynamically varying computing in the present more and more mobile computing.\textit{ Using computing under those extreme conditions
led to shocking and counter-intuitive experiences}
that can be more easily comprehended and accepted using parallels with the modern science~\cite{VeghModernParadigm:2019}.

Developing a new computing paradigm being able to provide a theoretical basis for state of the art cannot be postponed anymore. Based on that, one must develop different types of processors. As was admitted following the failure of  supercomputer Aurora'18: "\textit{Knights Hill} was canceled and instead be replaced by a "new platform and new microarchitecture specifically designed for exascale""~\cite{IntelDumpsXeonPhi:2017}. Similarly, we expect shortly to admit that building large-scale \gls{AI} systems is simply not possible based on the old paradigm and  architectural principles~\cite{DeepNeuralNetworkTraining:2016,VeghAIperformance:2020}.
The new architectures, however, require a new computing paradigm, that can give a proper reply to power consumption and performance issues of our present-day computing.

\subsection{Overview of the modern paradigm}
The new paradigm proposed here is based on fine distinctions in  some points, present also in the old paradigm. Those points, however, must be scrutinized individually,
whether and how long omissions can be made.  These points are:

\begin{itemize}
	\item consider that\textit{ not only one processor} (aka Central Processing Unit) exists, i.e.
	\begin{itemize}
		\item processing capability 
		is \textit{one of the resources} rather than a central singleton
		\item not necessarily \textit{the same processing unit} 
		is used to solve all parts of the problem
		\item a kind of redundancy (an easy method of replacing a flawed processing unit) through using virtual processing units is provided (mainly to \textit{increase the mean time between technical errors})
		\item instruction stream can be transferred to another processing unit~\cite{ARM:big.LITTLE:2011,Congy:CoreSpilling:2007}
		\item \textit{different processors can and must cooperate}  in solving a task, including direct data and control exchange between cores, communicating with each other, being able to set up ad-hoc assemblies for more efficient processing in a flexible way
		\item the large number of processors can used for \textit{replacing memory operations with using more processors}
		\item a core can outsource 
		the received task
	\end{itemize}
	\item misconception of segregated computer components is reinterpreted
	\begin{itemize}
		\item \textit{efficacy of using a vast number of processors
			is increased} by using multi-port memories (similar to~\cite{Cypress15})
		\item a "memory only" concept (somewhat similar to that in~\cite{ScratchpadMemory:2002}) is introduced (as opposed to the "registers only" concept), using \textit{purpose-oriented, optionally distributed, partly local, memory banks}
		\item principle of locality is introduced at hardware level, through introducing hierarchic buses
	\end{itemize}

	\item misconception of "sequential only" execution~\cite{BackusNeumannProgrammingStyle} is reinterpreted
	\begin{itemize}
		\item von Neumann required only "proper sequencing" for a single processing unit; this concept is \textit{extended} to several processing units
		\item tasks are broken into reasonably sized and logically interconnected fragments
		\item the "one-processor-one process" principle remains valid for task fragments, but not necessarily for the complete task
		\item fragments can be executed (at least partly) simultaneously if  both data dependence and hardware availability enables it (another kind of asynchronous computing~\cite{IBMAsynchronousAPI2017})
	\end{itemize}
	\item a closer hardware/software cooperation is elaborated
	\begin{itemize}
		\item hardware and software only exist together: the programmer works with virtual processors in the same sense as~\cite{ArpaciDusseau14-Book} uses these term and lets computing system to adapt itself to its task at run-time, through mapping  virtual processors to physical cores
		\item when a hardware has no duty, it can sleep ("does not exist", does not take power)
		\item the overwhelming part of the duties such as {synchronization, scheduling} of the OS are taken over by the hardware
		\item the compiler helps work of the processor with compile-time information and the processor can adapt (configure) itself to its task depending on the actual hardware availability
		\item strong support for multi-threading, resource sharing and low real-time latency is provided, at \gls{HW} level
		\item the internal latency of  large-scale systems is much reduced, while their performance is considerably enhanced
		\item task fragments shall be able to return control voluntarily without the intervention of  \gls{OS}, enabling to implement more effective and more simple operating systems
		\item the processor becomes "green": only working cores take power
	\end{itemize}
\end{itemize}

\subsection{Details of the concept}

We propose to work
at programming level with \textit{virtual processors} and to map them to physical cores at run-time, i.e., to \textit{let
	the computing system to adapt itself to its task}.
A major idea of \textit{EMPA} is to use \textit{quasi-thread (QT)} as atomic unit of processing,
that comprises both \textit{HW} (the physical core) and the \textit{SW} (the
code fragment running on the core). Its idea was derived with
having in mind the best features of both \textit{HW} core and \textit{SW} thread.
\textit{QT}s have "\textit{dual nature}"~\cite{VeghModernParadigm:2019}: in the \gls{HW} world of "classic computing"
they are represented as a 'core', in \gls{SW} world as a 'thread'. However, they are the same entity
in the sense of 'modern computing'.
We borrow the terms 'core' and 'thread' from conventional computing,
but in 'modern computing', they can actually exist only together in a time-limited way\footnote{Akin to dynamic variables on the stack: their lifetime is limited to the period when the HW and SW are appropriately connected. The physical memory is always there, but it is "stack memory" only when handled adequately by the HW/SW components.}. \textit{ EMPA is a new computing paradigm} which needs a new underlying architecture, \textit{rather than a  new kind of parallel processing}
running on a conventional architecture,
so \textit{it can be reasonably compared to terms and ideas used in conventional computing only in a minimal way};
although the new approach adapts many of its ideas and solutions from
'classic computing'.

One can break the executable task into reasonably sized and loosely dependent \gls{QT}s. (The \gls{QT}s can optionally be embedded into each other, akin to subroutines.) In \textit{ EMPA}, \textit{for every new \gls{QT} 	a new independent \gls{PU} is also implied, the internals (PC and part of registers) are set up properly},
and they execute their task independently\footnote{Although the idea of executing the single-thread task "in pieces" may look strange for the first moment, the same happens when the OS schedules/blocks a task.
	The key differences are that in EMPA \textit{not the same} processor is used, the \gls{EMPA} cuts the task into fragments in a  reasonable way (preventing issues like priority inversion~\cite{PriorityInversion:1993}). The \gls{QT}s can be processed \textit{at the same time} as long as their mathematical dependence \underline{and} the actual HW resource availability enable it.}
(but under the supervision of the processor comprising the cores).

In other words: \textit{we consider processing capacity as a resource} in the same sense as memory is considered as a storage resource. This approach enables the programmers to work with
virtual processors (mapped to physical \textit{PU}s by the computer at run-time)
and they can utilize the quick resource \textit{PU}s to replace utilizing the slow resource memory (say, renting a quick processor from a core pool can be competitive with saving and restoring registers in the slow memory, for example when making a subroutine call). The third primary idea is that \textit{PUs can cooperate}
in various ways, including data and control synchronization, as well as \textit{outsourcing part of the received job (received as an embedded \textit{QT})} to a helper core.
An obvious example is to outsource the housekeeping activity to a helper core: counting, addressing, comparing, can be done by a helper core, while the main calculation remains to the originally delegated core. As mapping  to physical cores occurs at run-time, (a function of actual HW availability)
the processor can avoid using (maybe temporarily) denied cores as well as to adapt the resource need (requested by the compiler) of the task to actual computing resource availability.

Processor has an additional control layer for organizing joint work of its cores. Cores have just a few extra communication signals and can execute both
conventional and so-called meta-instructions (for configuring their internal architecture).
A core executes a meta-instruction
in a co-processor style: when finding a meta-instruction, the core notifies
its processor which suspends conventional operation of the core, then controls executing
the meta-instruction (utilizing resources of the core, providing helper cores and handling connections between the cores as requested), then resumes core operation.

The processor needs to find the needed \textit{PUs} (cores), and its processing ability has to accommodate to the received task. Also, inside the processor, quickly, flexibly,
effectively, and inexpensively. \textit{A kind of `On demand' computing that works
	`As-a-Service'}. This task is not only for the processor: the complete computing system must participate, and for that goal, the complete computing stack must be rebuilt.

Behind the former attempts to optimize code execution inside the processor, there was no established theory, and they had only marginal effect because processor is working in real-time, it has not enough resources,
knowledge and time do discover those options entirely~\cite{WallLimitsOfILP:1993}.
In contrary, compiler can find out anything about enhancing performance but has no information about the actual run-time HW availability. Furthermore, it has no way to tell its findings to the processor. Processor has HW availability information but has to "reinvent the wheel" to enhance its performance; in real-time. In EMPA, compiler puts its findings in the executable code in form of meta-instructions
("configware"), and the actual core executes them with the assistance of the new control layer of the processor.
The processor can choose from those
options, considering the actual HW availability, in a style '\textbf{if} NeededNumberOfResourcesAvalable \textbf{then} Method1 \textbf{else} Method2', maybe nested one into another.

\subsection{Some advantages of EMPA}
The approach results in several considerable advantages,
but the page limit enables us to mention just a few of them.
\begin{itemize}
	\item as a new \textit{QT} receives a new \gls{PU}, there is no need to save/restore registers and return address
	(less memory utilization and less instruction cycles)
	\item OS can receive its \gls{PU}, initialized
	in kernel mode and can promptly (i.e., without the need of context change) service the requests from the requestor core
	\item for resource sharing,  a \gls{PU} can be temporarily
	delegated to protect the critical section; the next call to
	run the code fragment with the same offset shall be delayed (by the processor) until processing by the first \gls{PU} terminates
	\item processor can natively accommodate to the variable
	need of parallelization
	\item out-of-use cores are waiting in low energy consumption mode
	\item hierarchic core-to-core communication greatly increases memory throughput
	\item asynchronous-style computing~\cite{AsynchronParadigm:2013} largely reduces loss stemming from the gap~\cite{NinjaPerformanceGap:2015:CACM} between speeds of processor and memory
	\item \textit{principle of locality can be applied inside the processor}:  direct core-to-core connection (more dynamic than in~\cite{CooperativeComputing2015}) greatly enhances efficacy in large systems~\cite{TaihulightHPCG:2018}
	\item the communication/computation ratio, defining decisively efficiency~\cite{ScalingParallel:1993, VeghHowMany:2020,VeghAIperformance:2020},
	is reduced considerably
	\item \gls{QT}s thread-like feature akin to $fork()$ and hierarchic buses change the dependence of the time of creating many threads
	on the number of cores from linear to logarithmic
	(enables to build exascale supercomputers)
	\item inter-core communication can be organized in some sense similar to \gls{LAN}s of computer networking.
	For cooperating,
	cores can prefer cores in their topological proximity
\end{itemize}

\section{How to implement EMPA}\label{sec:Implementation}

 The best starting point to understand  implementation of the \gls{EMPA} principles is conventional many-core processors. Present electronic technology made kilo-core processors available~\cite{KilocoreChip:2017,PEZY2048cores:2017}, in a very inexpensive way and in immediate proximity of each other, in this way making the computing elements a "free resource"~\cite{SpiNNaker:2013}.
 Principles of \gls{SPA}, however, enable us to use them in a rather ineffective way~\cite{HillMulticoreAmdahl2008}.

\begin{figure*}
\maxsizebox{\textwidth}{!}
	{
		\includegraphics[width=\textwidth]{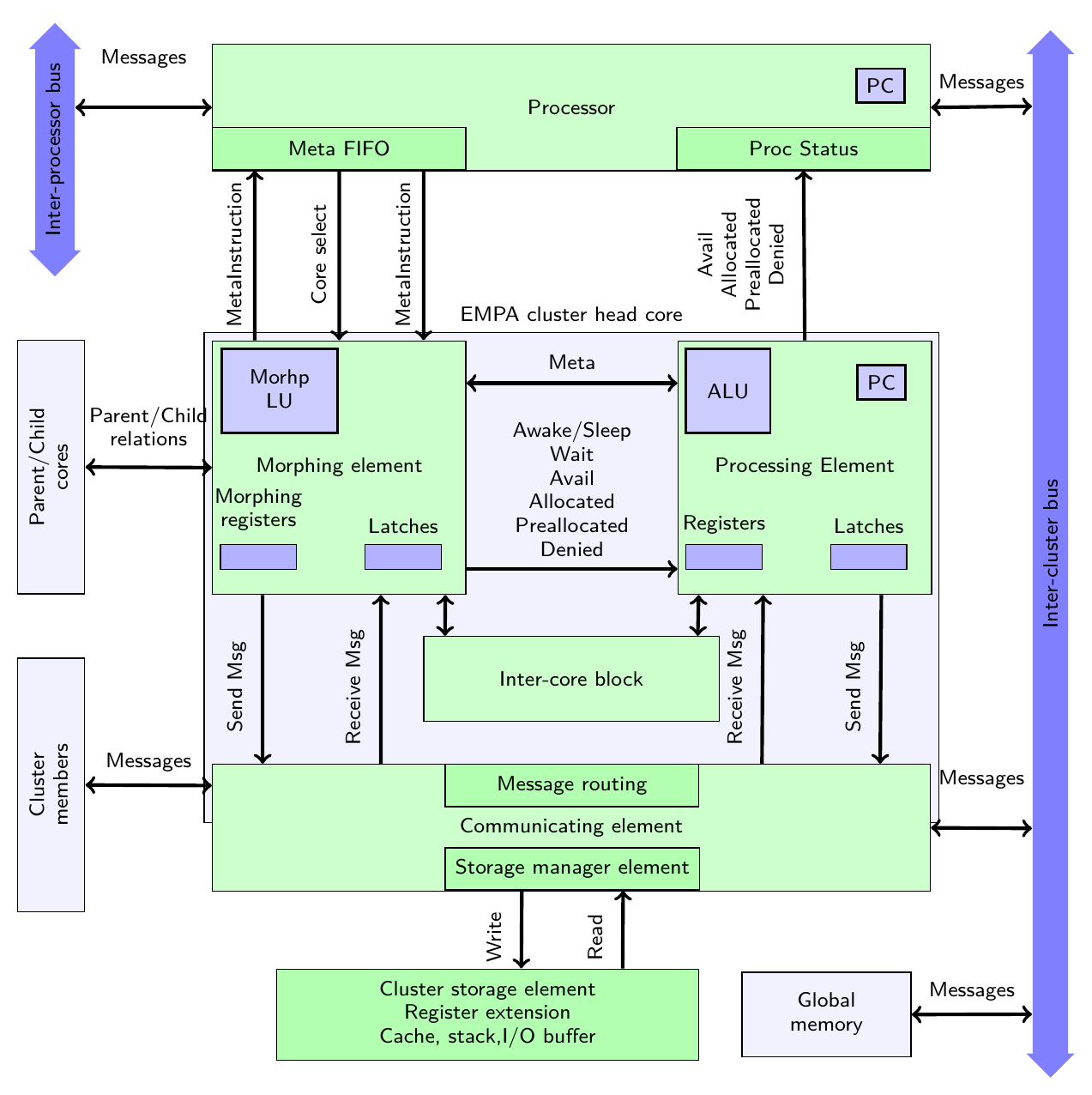}
	}

	\caption{The logical overview of the EMPA-based computing.\label{fig:EMPAoverview}}
\end{figure*}

Given that true parallelism
cannot be achieved (working with the components anyhow
needs time and synchronization via signals and/or messages,
the question is only the time resolution), \textit{\gls{EMPA} targets an enhanced and synchronized parallelized sequential processing based on using many cooperating processors}.
The implementation uses variable granularity and as much truly parallel portions as possible.
However, \textit{focus is on the optimization of the operation, rather than providing some new kind of parallelization}. The ideas of cooperation comprise job outsourcing, sharing different resources and providing
specialized many-core computing primitives in addition to the
single-processor instructions.

In this way \textit{\gls{EMPA} is an extension of \gls{SPA}}: conventional computing is considered
consisting of a single non-granulated thread,
where (mostly) \gls{SW} imitates the required illusion
of granulating and synchronizing code fragments.
Mainly because of this, many of components have a name and/or functionality
familiar from conventional computing.
However, there are subtle details that are different
from those of conventional computing.
Furthermore, we consider  the \textit{computing process as a whole} to be the subject of optimization rather than the segregated components individually.

In \gls{SPA}, there is only one active element,
the \gls{CPU}. The rest of components of the system serves requests from the \gls{CPU}
in a passive way. As the \gls{EMPA} wants to \textit{extend} conventional computing, rather than to \textit{replace} it, its operating principle is somewhat similar to the conventional one,
with important differences in some key points. Fig.~\ref{fig:EMPAoverview} provides an overview of the operating principle and major components of \gls{EMPA}.
We follow hints by Amdahl :
"\textit{general purpose computers with a generalized interconnection
	of memories or as specialized computers with geometrically related memory interconnections and
	controlled by one or more instruction streams}"~\cite{AmdahlSingleProcessor67}.

\subsection{The core}
An \gls{EMPA} core of course comprises an \gls{EPE}. Furthermore, it addresses two key deficiencies of conventional computing : the inflexibility of computing architecture by \gls{EME}, and the lack of autonomous communication by \gls{ECE}.
Notice the important difference to conventional computing: \textit{the next instruction can be taken either from memory pointed out by the instruction pointer or from the Meta FIFO}.

\subsubsection{The Processing Element}

The \gls{EPE} receives an address, fetches the instruction (if needed, also its operands).
If the fetched instruction is a meta-instruction,
\gls{EPE} sets its 'Meta' signal (changes to 'Morphing' regime) for the \gls{EME}
and waits (suspends processing instructions) until the \gls{EME} clears that signal.

\subsubsection{The Morphing Element}
When \gls{EPE} sets the  'Meta' signal,
\gls{EME} comes into play. Since the instruction and its operands are available, it attempts to process the received meta-instruction.
However, the meta-instruction refers to resources handled by the processor. At processor level, order of execution of meta-instructions depends on their priority.  Meta-instructions, however,
may handle the 'Wait' or the core signal correspondingly.
Notice that the idea is different from configurable spatial accelerator~\cite{patent:20180189231,IntelDataflowPatent:2018}: the needed configuration is assembled ad-hoc, rather than chosen from a list of preconfigured assemblies.
\index{signal!Meta}\index{signal!Wait}

Unlike in \gls{SPA}, communication is a native feature of \gls{EMPA} cores
and it is implemented by \gls{ECE}.
The core assembles the message content (including addresses), then after setting a signal, the message is
routed to its destination, without involving a computing element and without any respect to where the destination is.
The message finds its path to its destination autonomously, using \gls{EMPA}'s  hierarchic bus system and  \gls{ECE}s of the fellow cores, taking the shortest (in terms of transfer time) path.
Sending messages is transparent for both programmer and \gls{EPE}.

\subsubsection{The Storage Management Element}
\gls{ESME} is implemented only in cluster head cores, and its task is to
manage storage-related messages passing through \gls{ECE}.
It has the functionality (among others) similar to that of  memory management unit and cache controller in conventional computing.

\subsection{Executing the code}
\subsubsection{The quasi-threads}

Code (here it means a reasonably sized sequence of instructions) execution begins with 'hiring' a core: the cores
by default are in a 'core pool', in low energy consumption mode.
\index{core!hired}
\index{core pool}
The 'hiring core' asks for a helper core from the processor. If no cores are
available at that moment, the processor sets the 'Wait' signal for the requester core and keeps the request pending.
At a later time, processor can serve this pending request with a 'reprocessed' core.

Notice that the idea is quite different from the idea of
e\textbf{X}plicit \textbf{M}ulti\textbf{T}hreading~\cite{VishkinHome2007,SpawnJoinArchitectureVishkin:1998}.
Although they share some ideas such as the need for fine-grained multi-threaded programming
model and architectural support for concurrently executing multiple contexts on-chip,
unlike XMTs, \gls{QT}s embody not simply mapping the idea of multi-threading to \gls{HW} level.
They are based on a completely unconventional computing paradigm; the \gls{QT}s can be nested.

\textit{This operating principle also means that the code fragment and the active core
exist only together, and this combination (called \gls{QT}) has a lifetime.}
\index{quasi-thread}\index{QT}
The principle of the implementation is akin to that of the 'dynamic variable'. \gls{EMPA} hires a core for executing a well-defined code fragment, and only for the period between creating and terminating a \gls{QT}. In two different executions, the same code fraction may run on different physical cores.

\subsubsection{The process of code execution}

When a new task fragment appears, an \gls{EMPA} processor must provide a new computing resource for that task fragment (a new register file is available). Since the executing core is 'hired' only for the period of executing a specific code fragment, it must be returned to core pool when execution of the task fragment terminates. The 'hired' \gls{PU}
is working on behalf of the 'hiring' core, so it must have
the essential information needed for performing the task. The core-to-core register messages provide a way to transfer register contents from the parent core to the child core.

Beginning the execution of an instruction sets the signal 'Meta', i.e.
\index{signal!Meta'}
selects either \gls{EPE} or \gls{EME} for the execution,
and that element executes the requested action.
The core repeats the process until the 'hired' core finds
and 'end of code fragment' code.
Notice the difference to conventional computing: processing of the task does not terminate; only the core is put back into 'core pool' as at the moment it is not anymore needed.


When 'hired' core becomes available,  processing continues with fetching an instruction by the 'hired' core.
\index{core!hired}
For this, the core sends a message with the address of the location of the instruction. The requested memory content arrives at the core in a reply message logically from the addressed memory, but the \gls{ESME} typically intercepts the action.
\index{message!Memory}
The process is similar to that in the conventional computing. However, here the memory requests to send a reply to the request when it finds the requested contents. Different local memories, such as addressable cache, can also be handled.
 Notice also that the system uses complete messages (rather than simple signals with the address); this makes the way of accessing some content independent from its location, although it needs location-dependent time.

Of course, 'hiring' core wants to get back some results from the 'hired' core. When starting a new \gls{QT},
'hiring' core also defines, with sending a mask, which register contents the hired core shall send back. In this case,  synchronization is a serious issue: parent core utilizes its registers for its task, so it is not allowed to overwrite any of its registers without an explicit request from the parent.
Because of this, when a child terminates, it writes the expected register contents to the latch storage of the parent, then it may go back to the 'core pool'.
When the parent core reaches the point where it needs the register contents received from its child, it explicitly asks to clone the required contents from the latches to its corresponding register(s). It is the parent's responsibility to issue this command at such a time when no accidental register overwriting can take place.

Notice that beginning execution of a new code fragment
needs more resources, while terminating it frees some resources. Because of this, terminating a \gls{QT}
has a higher priority than creating one.
This policy, combined with that the cores are able
to wait until their processor can provide the requested amount of resources, prevents "eating up" the computing resources
when the task (comprising virtually an infinite number of \gls{QT}s) execution begins.

\subsubsection{Compatibility with conventional computing}

Conventional code shall run
on an \gls{EMPA} processor (as an implicitly created \gls{QT}). However, that code can only use a single core, since it contains no meta-instructions to create more \gls{QT}s.
This feature enables us to mix \gls{EMPA}-aware code with conventional code, and (among others) enables us to use the plethora of standard libraries without rewriting that code.

\subsubsection{Synchronizing the cooperation}

The cores execute their instruction sequences independently,
but their operation must be synchronized at several points.
Their initial synchronization is trivial: processing
begins when the 'hired' core received all theits required operands (including instruction pointer, core state, initial register contents, mask of registers the contents of which the hiring core requests to return).
The final synchronization on the side of the 'hired' core
is simple: the core simply sends the contents of the registers as was requested at the beginning of the execution of the code fragment.

\begin{figure}
	\maxsizebox{\columnwidth}{!}
	{
		\includegraphics{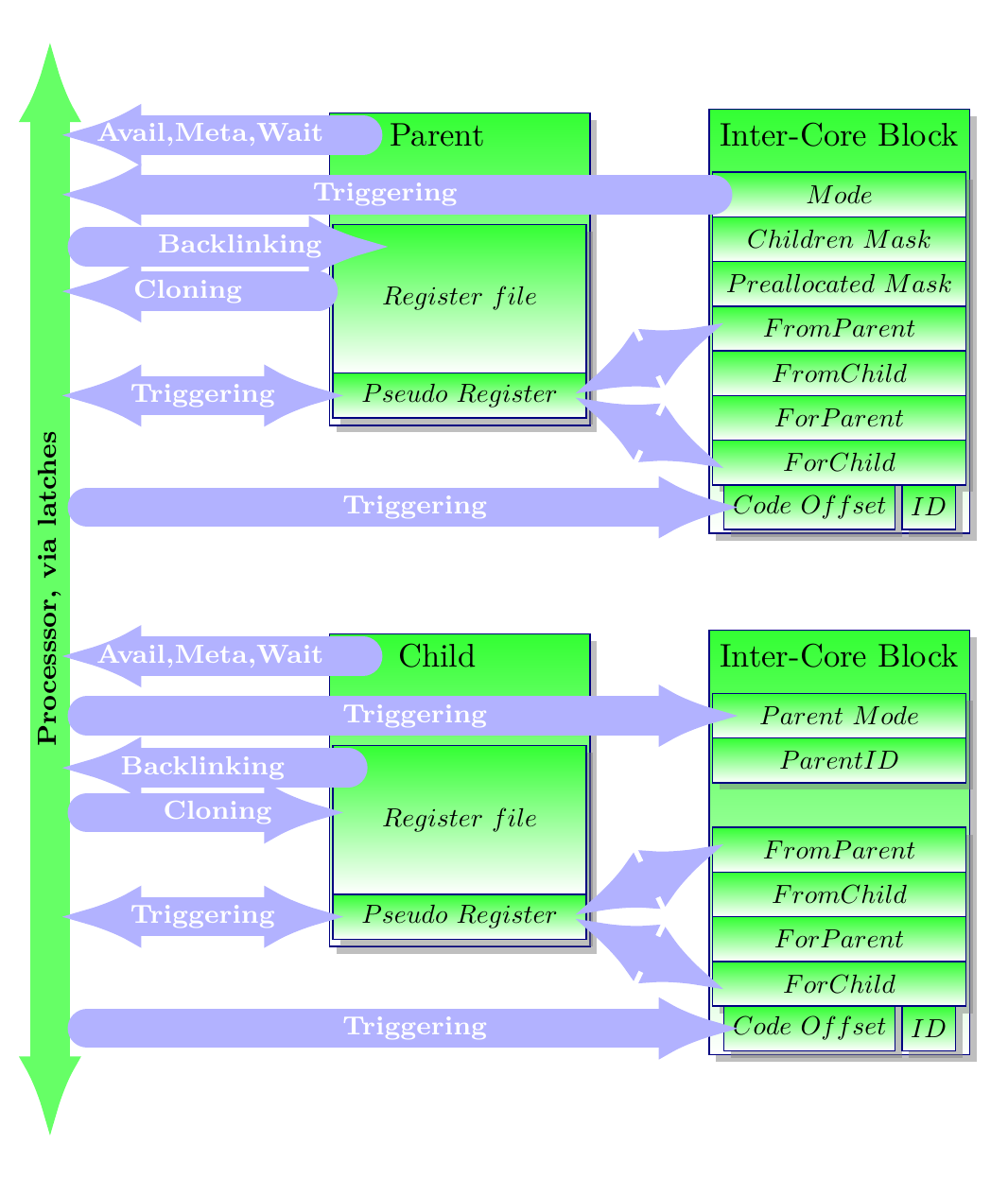}
	}

	\caption{Implementing the parent-child relationships: registers and operations of the \gls{EICB}\label{fig:EMPAParentChild}}
\end{figure}

On the side of the 'hiring' core, the case is much more complex. The 'hiring' core may wait for the termination of the code fragment running on the 'hired' core, or maybe it is in the middle of its processing. In the former case, a simple waiting
until the message arrives is sufficient, but in the latter case, receiving some new register contents in some inopportune time
would destroy its processing. Because of this, the register contents from the 'hired' core are stored temporarily in latch registers, and they are copied to the corresponding registers of the 'hiring' core only when the 'hiring' core
requests so explicitly. Fig.~\ref{fig:EMPAParentChild} attempts to illustrate the complex cooperation between the \gls{EMPA} components.

\subsection{Organizing 'ad hoc' structures}
\gls{EME} can 'morph' the internal architecture
of the \gls{EMPA} processor, as required by the actual task (fragment).
\gls{EMPA} uses the principle of creating 'parent-child' (rather than
'Master-Slave')  relation between its cores. The 'hiring' core
becomes the parent, and the 'hired' core becomes the child.
A child has only one parent, but parents can have any number of children. Children can become parents in some next phase of execution; in this way, several 'generations' can cooperate. This principle provides a dynamic processing capacity for different tasks (in different phases of execution). The 'parent-child' relations simply mean storing addressing information,
in the case of children combined with concluding the address from the 'hot' bits of a mask.

As 'the parents are responsible for their children',
parents cannot terminate their code execution until
all their children returned the result of the code fragment that they delegated for them. This method enables parents also to trust in their children: when they delegate some fragment of their code to their children, they can assume that that code fragment is (logically) executed. It is the task of the
compiler to provide the required dependence information,
how those code fragments can be synchronized.

This fundamental cooperation method enables the purest form of delegating code to existing (and available) cores.
In this way, all available processing capacity can be used, while only the actually used cores need energy supply (and dissipate). \textit{Despite its simplicity, this feature enables us to make subroutine calls without needing to save/restore contents through memory and to implement mutexes working thousands of times quicker than in conventional computing}.

\subsection{The processor}
The processor comprises many physical \gls{EMPA} cores. The \gls{EMPA} processor appears
in the role of a 'manager' rather than a number-crunching unit, it only manages its resources.


Although individual cores initiate  meta-instructions,
their synchronized operation  requires the assistance of the processor.
Meta-instructions received by \gls{EMPA} cores are written first (without authorization) in a priority-ordered queue (Meta FIFO) in the processor,
\index{meta-instruction FIFO}
so the processor can always read and execute only the highest priority meta-instruction (a core can have at most one active meta-instruction).

\subsection{Clustering of the cores}

The idea of arranging \gls{EMPA} cores to form clusters
is somewhat similar to that of CNNs~\cite{CellularNeuralNetworks93}.
In computing technology, one of the most severe limitations is given by 
internal wiring, both for internal signal propagation time and area
occupied on the chip~\cite{LimitsOfLimits2014}.
In conventional architectures, cores are physically arranged to form a 2-dimensional rectangular grid matrix. Because of  \gls{SPA}, there should not be any connection between segregated cores, so the inter-core area is only used by some kind of internal interconnection networks or another wiring.

\begin{figure}
	\maxsizebox{\columnwidth}{!}
	{
		\includegraphics[width=\textwidth]{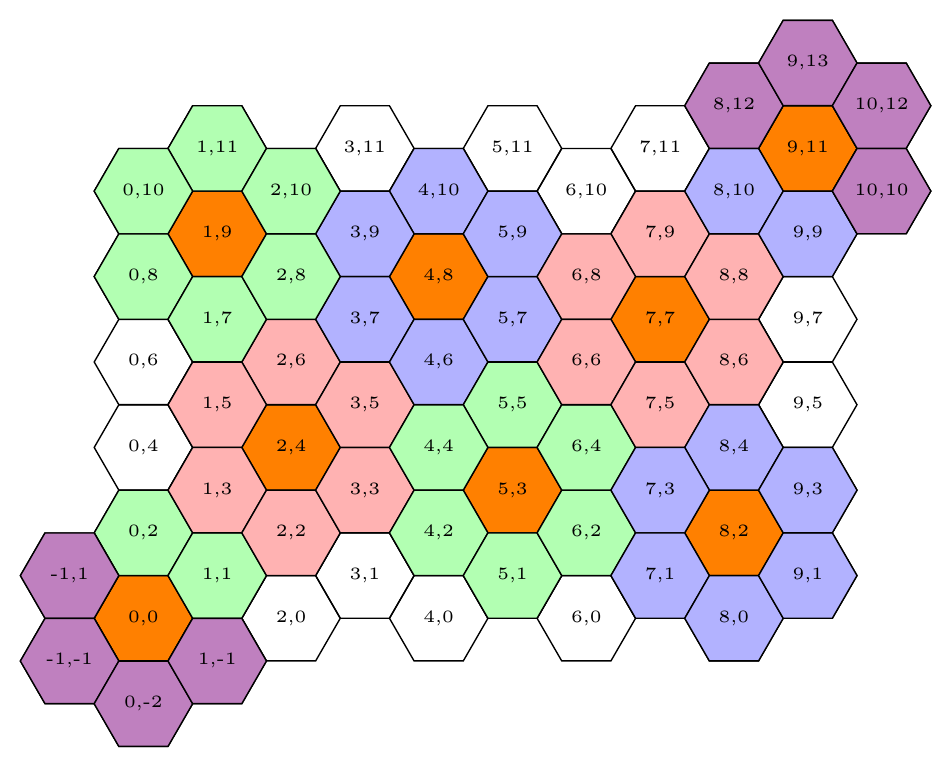}
	}

	\caption{The (logically) hexagonal arrangement (internal clustering) of EMPA cores in the EMPA processor\label{fig:Hexagonal}}
\end{figure}

In \gls{EMPA} processors,
even-numbered columns in the grid are shifted up by a
half grid position. In this way cores are arranged in a way that they
have common boundaries with cores in their neighboring columns.
In addition to these neighboring cores, cores have (up to two) neighbors in their column,
with altogether up to six immediate neighbors, with common boundaries.
This method of positioning also means that
cores, logically, can be arranged to form a hexagonal grid, as shown in Fig.~\ref{fig:Hexagonal}.
Cores \textit{physically} have a rectangular shape with joint boundaries with their neighbors,
so  \textit{logically} the cores form a hexagonal grid.
This positioning enables
to form "clusters" of cores, forming a "flower": an orange
\textit{ovary} (the cluster head) and six \textit{petal}s
(the leaf cores of the cluster, the members).

Between cores arranged in this way also neighborhood size can be interpreted similarly to the case of cellular computing.
\index{neighborhood size}
Based on neighborhood of size $r=1$ (that means that cores have precisely one common boundary),
a cluster comprising up to six cores (cluster members) can be formed, with the orange cell (of size $r=0$, the cluster head)
in the middle.
Cluster members have shared boundaries with their immediate neighbors, including their cluster head.
These cores define the
external boundary of the cluster (the "flower").
Cores within this external boundary are
the "ordinary members" of the cluster, and the one
in the central position is the \textit{head of the cluster}.

There are also "corresponding members" (of size $r=2$): cores having at least one common boundary with one of the "ordinary members" of the cluster head in question.
"Corresponding members" may or may not have their cluster head, but have a common boundary with the "ordinary members".
The white cells in the figure represent "external members" (also of size $r=2$): they have at least one common boundary with an "ordinary member", like the "corresponding members", but unlike the "corresponding members" they do not have their cluster head. Also, there are some "phantom members" (see the violet petals in the figure) around the square edges in the figure: they have a
cluster head and the corresponding cluster address,
but (as they are physically not implemented in the square grid of cores during the manufacturing process)
they do not exist physically.

That means: a cluster has one core as "cluster head"; up to six "ordinary members", and up to twelve "corresponding members";
i.e., an "extended cluster" can also be formed, comprising up to 1+6+12 members. Notice that around the edge of the square grid "external members" can be in the position of the "corresponding members", but the upper limit of the total number of  members in an extended cluster does not change. Interpreting members of size $r>=2$ has no practical importance. \textit{The cores with $r<=2$ have a direct communication mechanism}.

\begin{figure}
	\maxsizebox{\columnwidth}{!}
	{
		\includegraphics{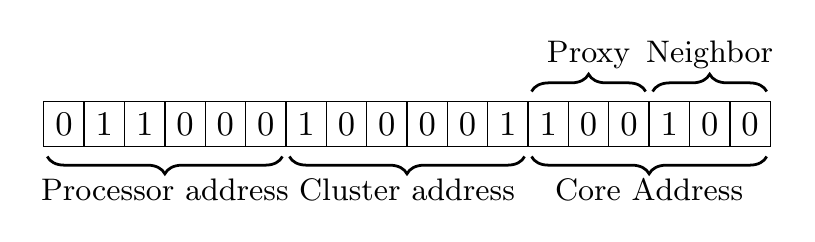}
	}

	\caption{Implementing the hierarchical cluster-based addressing bit fields of the cores of EMPA processors. A cluster address is globally unique.\label{fig:ClusterAddressing}}
\end{figure}

The addressing system must provide support for all those addressing modes. The cluster addressing is of central importance because of the topology of cores: \textit{the cores having common boundary surely do not need a bus between the neighboring cores}.
The addressing must support the goal to keep the messages inside the cluster, if possible. Messages from/to outside the cluster are received/sent by the cluster head. The rest of the messages are sent directly or with using a proxy
to their final destination. To implement that goal, \gls{EMPA} processors use the addressing scheme shown in Fig.~\ref{fig:ClusterAddressing}. Notice that the proposed addressing system a network logical address can be directly
(and transparently) mapped to the ID and vice versa.

In \gls{EMPA}, cluster addressing carries also  topological information, partly relies on relative topological positions,
and enables to introduce different classes of relationship between cells. As mentioned, cluster head cores have a physically distinct role  (In this sense, they can also be a "fat" core) and enables us to introducecluster addressing for members of the extended clusters.
Only cluster head cores have an immediate global memory access (considerably reducing the need
for wiring). The cores being in  neighborhood of size $r=1$
can access memory through their cluster head. These cores can also be used as a proxy for cores in neighborhood of size $r=2$. The latter also enables to replace a denied cluster head core.

In \gls{SPA}, the grid and linear addressing are purely logical ones, which use absolute addresses known at compile time.
Similarly to computer networks, \gls{EMPA} cores have
(closely related through the cluster architecture)
both logical and physical addresses, enabling autonomous (computing-unrelated) communication and virtual addressing.

\subsection{The compiler}
Compiler plays a significant role in the \gls{EMPA}.
It should discover all possibilities of cooperation,
especially the ones that become newly available
with the philosophy that \textit{the appearance of a new task
is attached with the appearance of new computing resource, with a
new register file}. Because at the time of compilation
actual \gls{HW} availability cannot be known,
code for different scenarios must be prepared and put
in the object code.

The philosophy of coding must be drastically changed.
Given that, with outsourcing, a new computing facility appears, and processor assures the proper synchronization and data transfer, there is no need to store/restore return address and save/restore data in the registers, leading to less memory traffic, and quicker execution time.

The object code is essentially unchanged, except that some fragments (the \gls{QT}s) are bracketed by meta-instructions. The \gls{QT}s  can be nested (i.e., meta-instructions are inserted into conventional code). One can consider that \gls{QT}s represent a kind of atomic macros
which have some input and output register contents but do not need processing capacity from the
actual core.

\section{The new features the EMPA offers\label{sec:NewFeatures}}

Although \gls{EMPA} does not want to address \textit{all} of the challenges of computing,
it addresses many of them (and leaves the door open for addressing further challenges). Due to lack of space, code examples, comparisons, and evaluations, based on the loosely-timed SystemC simulation~\cite{VeghEMPAthY86:2016}, are left for simulator documentation and the early published version~\cite{RenewingComputingVegh:2018}.

\subsection{Architectural aspects}

Notice that ad hoc assemblies  consider both
current state of the cores, and also their 'Denied' signal.
\index{signal!Denied}
That is, the flawed (or just temporarily overheated) cores are not used, significantly increasing the mean time between machine failures.
Also, notice that this approach enables using 'hot swap' cores,
in this way providing dynamic, connected systems
(the addressing is universal, and the information is
delivered by messages; it takes time, but possible),
as well as \textit{to deliver the code to the data}: the physical cores
can be located in the proximity of the 'big data' storage,
the instruction is delivered to the place, and only the
processed, needed result is to be transported back.

\subsubsection{Virtualization at HW level}
In \gls{EMPA} no absolute processor levels are utilized:
virtual processors seen by the programmer are mapped 'on the fly' to physical core by the \gls{EMPA} processor.
Physical cores have a 'denied' state that can be set permanently (like fabrication yield) or temporarily (like overheating), in which case the core
will not be used to map a virtual core to it.
When combined with a proper self-diagnostic system, this feature prevents
extensive systems to fail because of a failing core.
The processor has the right and possibility to replace a physical core
any time with another one.

\subsubsection{Redundancy}
Huge masses (literally millions/billions) of silicon-based elements are deployed in all systems.
As a consequence, the components showing a tolerable error rate in "normal" systems,
but (purely due to the high number of components)  need special care in the case of
large-scale systems~\cite{CosmicRaysSupercomputers:2018}.

The usual engineering practice is to rely on the high reliability of the components. The fault-tolerant systems require particular technologies, typically majority voting, but they are also based on the same single high-reliability components.

\subsubsection{Reduced power consumption}

The operating principle of a processor is based on the assumption
that processors are working continuously, executing instructions one after the other, as their control unit defines the required sequencing.
Because of this principle, in the \gls{OS} an 'idle' task is needed.
In \gls{EMPA}, cores can return control voluntarily, enabling most of the cores to stay in an 'idle' state.

\subsection{Attacking the memory wall}
The 'memory wall' is known as the 'von Neumann' bottleneck of computing, especially after that memory access time became hundreds of times slower than processing time. Although 'register only' processing and cache memories can seriously mitigate its effect, in the case of large systems the 'sparse' calculations that poorly use the cache, show up orders of magnitude worse computing efficacy, i.e., further improvement in
using the memory is of utmost importance.

\subsubsection{Register-to-register transfer }
The idea of immediate register-to-register transfer~\cite{CooperativeComputing2015} seriously can increase the performance of real-life tasks~\cite{TaihulightHPCG:2018}. In \gls{EMPA}, the idea is used in combination with the flexibility via using virtual cores, multiple register arrays via children.

\subsubsection{Subroutine call without stack }
In  \gls{SPA}, a subroutine call requires
to save/restore the return address and (at least part of) the register file; unfortunately, one can use only the main memory for that temporary storage.
In \gls{EMPA}, for executing subroutine code, another \gls{PU} is provided. Because of this solution, \gls{HW} can remember (in a nested way) the return address. Furthermore, working area is provided by the register file of the 'hired' core. Given that a register-to-register transfer is provided, code execution can be hundreds of times quicker.
With proper organization, hiring and hired cores can also run partly parallel.

\subsubsection{Interrupt and systems calls without context switching}
Given that interrupts and \gls{OS} service calls can be considered as special service calls, where also context switching is needed, using a prepared (waiting in kernel mode) core can service the request thousands of times quicker.
Event, interrupts can be serviced \textit{without interrupting} the running process.

\subsubsection{Resource sharing without scheduling}
For multitasking, only the \gls{OS} can provide exclusive access to some resource (as in \gls{SPA}, no other processor/task exists). \gls{EMPA} offers a simple, elegant, and quick solution: it can delegate a \gls{QT} for the task of guarding a critical section, and all tasks issue a conditional subroutine call to the code guarded by that \gls{QT}. All but the first requester \gls{QT} must wait (but are scheduled automatically by the processor), and after servicing all requests, the delegated core is put back to the pool.
Since the compiler creates reasonably sized code fragments, cases leading to priority inversion~\cite{PriorityInversion:1993}
cannot happen, so no specialized protocols are needed in the \gls{OS}: the orchestrated work in \gls{EMPA} prevents those issues.

\subsection{Attacking the communication wall}
In \gls{SPA}, communication is not natively present (no other processor exists); it must be performed and synchronized using \gls{I/O} instructions and \gls{OS} operations, in payload processing time;
resulting in performing a severe amount of non-payload instructions.

\subsubsection{Decreasing the internal latency}
When using interconnected cores, \gls{ECE} can take over most of the non-payload duties,
enabling to decrease the sequential-only portions of the task that decisively define communication/computation ratio~\cite{ScalingParallel:1993}; a significant point when developing large scale computing systems~\cite{VeghHowMany:2020} or using \gls{AI}-type workloads~\cite{VeghAIperformance:2020}.

\subsubsection{Hierarchic (local) communication}
Using temporally or spatially local memory accesses can increase the efficiency dozens of times.
Similarly, providing 'interconnection cache' for the \gls{EMPA} processor can result in considerable improvement in final efficiency of the system. As computing tasks change their state between 'computing bound' and 'communication bound' dynamically, this solution mitigates both limiting factors as much as possible.

\subsubsection{Fully asynchronous operation}
As von Neumann only required a 'proper sequencing' of instructions, and having less 'idle' times during core operation appears as performance increase, asynchronous operation (i.e., turning all components to active) can considerably contribute to more effective (i.e., comprising fewer losses) operation.

\section{Summary}

In computing, the incremental development methods face more and more difficulties, because of the drastic changes both in technology and utilization. The final reason, as has been suspected by many researchers, is the computing paradigm reflecting a 70-year old state of the art.
Computing needs renewal~\cite{RenewingComputingVegh:2018}
and rebooting.
As a first step, the validity of restrictions was scrutinized. It was presented that \textit{it is not a necessary condition that the same computer} solves all the tasks: von Neumann only required a "proper sequencing" in executing machine instructions.
This requirement can be satisfied in a much better way via using the presently available many "free" processors. That way requires an entirely different thinking (and component base) and offers real advantages. We can implement the introduced new paradigm by putting the presently available technology solutions along with different principles
that approach offers considerable advantages.


\end{document}